# Application of Data Mining to Network Intrusion Detection: Classifier Selection Model

Huy Anh Nguyen and Deokjai Choi

Chonnam National University, Computer Science Department,
300 Yongbong-dong, Buk-ku,
Gwangju 500-757, Korea
anhhuy@gmail.com, dchoi@chonnam.ac.kr

**Abstract.** As network attacks have increased in number and severity over the past few years, intrusion detection system (IDS) is increasingly becoming a critical component to secure the network. Due to large volumes of security audit data as well as complex and dynamic properties of intrusion behaviors, optimizing performance of IDS becomes an important open problem that is receiving more and more attention from the research community. The uncertainty to explore if certain algorithms perform better for certain attack classes constitutes the motivation for the reported herein. In this paper, we evaluate performance of a comprehensive set of classifier algorithms using KDD99 dataset. Based on evaluation results, best algorithms for each attack category is chosen and two classifier algorithm selection models are proposed. The simulation result comparison indicates that noticeable performance improvement and real-time intrusion detection can be achieved as we apply the proposed models to detect different kinds of network attacks.

## 1 Introduction

In the era of information society, computer networks and their related applications are becoming more and more popular, so does the potential thread to the global information infrastructure to increase. To defend against various cyber attacks and computer viruses, lots of computer security techniques have been intensively studied in the last decade, namely cryptography, firewalls, anomaly and intrusion detection … Among them, network intrusion detection (NID) has been considered to be one of the most promising methods for defending complex and dynamic intrusion behaviors.

Intrusion detection techniques using data mining have attracted more and more interests in recent years. As an important application area of data mining, they aim to meliorate the great burden of analyzing huge volumes of audit data and realizing performance optimization of detection rules. Different researchers propose different algorithms in different categories, from Bayesian approaches [8] to decision trees [9, 10], from rule based models [3] to functions studying [11]. The detection efficiencies therefore are becoming better and better than ever before.

However, to the best of our knowledge, never before has it existed a considerable comparison among these classification methods to pick out the best ones that suite the job of intrusion detection. A literature survey that was done by us also indicates a fact that, for intrusion detection, most researchers employed a *single algorithm* to detect *multiple attack categories* with dismal performance in some cases. Report results suggest that much detection performance improvement is possible. In light of the widely-held belief that attack execution dynamics and signatures show substantial variation from one attack category to another, identifying attack category specific algorithms offers a promising research direction for improving intrusion detection performance.

In this paper, a comprehensive set of classifier algorithms will be evaluated on the KDD dataset [2]. We will try to detect attacks on the four attack categories: Probe (information gathering), DoS (deny of service), U2R (user to root), R2L (remote to local). These four attacks have distinct unique execution dynamics and signatures, which motivates us to explore if in fact certain, but not all, detection algorithms are likely to demonstrate superior performance for a given attack category. And from the performance comparison result of the classifiers, we will hereby propose the model for classifier algorithm selection regarding the best performing algorithms for each attack category.

The remainder of this paper is organized as follows. We make a quick and up-to-date literature survey on attempts for designing intrusion detection systems using the KDD dataset in Section 2. Section 3 will detail about our simulation study (classifiers, evaluation setup and performance comparison). Two models for algorithm selection will be proposed in Section 4. Section 5 is the performance comparisons to prove the effectiveness of our models; implementing issues will also be discussed here. Finally, Section 6 will conclude our study and discuss the future works.

## 2   Related Works on KDD99 Dataset

Agarwal and Joshi [4] proposed a two-stage general-to-specific framework for learning a rule-based model (PNrule) to learn classifier models on a data set that has widely different class distributions in the training data. The PNrule technique was evaluated on the KDD testing dataset which contained many new R2L attacks that were not presented in the KDD dataset. The proposed model was considered well-performed with the TP (*true positive*) rate at 96.9% for DoS, 73.2% for Probe, 6.6% for U2R and 10.7% for R2L attacks. FP (*false positive*) rate was generated at a level of less than 10% for all attack categories except for U2R: an unacceptably high level of 89.5% FP was reported.

Yeung and Chow [5] proposed a novelty detection approach using no-parametic density estimation based on Parzen-window estimators with Gaussian kernels to build an intrusion detection system using normal data only. This novelty detection approach was employed to detect attack categories in the KDD dataset. The technique has surprisingly good reported results: 96.71% of DoS, 99.17% of Probe, 93.57% of U2R and 31.17% of R2L respectively. However, due to the fact that no FP was reported by the authors and a nearly impossible detection rate [13] of 93.57% of U2R category, we really have to question the authentic of the reported numbers.

In 2006, Xin Xu et al. [6] presented a framework for adaptive intrusion detection based on machine learning. Multi-class Support Vector Machines (SVMs) is applied to classifier construction in IDSs and the performance of SVMs is evaluated on the KDD99 dataset. Promising results were given: 76.7%, 81.2%, 21.4% and 11.2% detection rate for DoS, Probe, U2R, and R2L respectively while FP is maintained at the relatively low level of average 0.6% for the four categories. However, this study can only use a very small set of data (10,000 randomly sampled records) comparing to the huge original dataset (5 million audit records). Therefore it is difficult to convince a strict audience about the effectiveness of the method.

Yang Li and Li Guo [7] though realize the deficiencies of KDD dataset, developed a supervised network intrusion detection method based on Transductive Confidence Machines for K-Nearest Neighbors (TCM-KNN) machine learning algorithm and active learning based training data selection method. The new method is evaluated on a subset of KDD dataset by random sampling 49,402 audit records for the training phase and 12,350 records for the testing phase. An average TP of 99.6% and FP of 0.1% was reported but no further information about the exact detection rate of each attack categories was presented by the authors.

Literature survey shows that, for all practical purposes, most researchers applied a single algorithm to address all four major attack categories with dismal performance in many cases. This leads to our belief that different algorithms will perform differently on different attack categories, and this also is the motivation of our study.

## 3  Empirical Study

In order to verify the effectiveness of different classifiers algorithms for the field of intrusion detection, we will use the KDD99 dataset to make relevant experiments step-by-step. Firstly, we build the experiment evaluation environment with major steps: environment setup, data preprocessing, choosing the data mining software. Secondly, we select a comprehensive set of most popular classifier algorithms, ten distinct widely used classifier algorithms were selected so that they represent a wide variety of fields: Bayesian approaches, decision trees, rule based models and function studying and lazy functions. An overview of how specific values of these algorithms were identified as well as their detection performance will be given. Finally, we come up with the performance comparison between the ten selected classifiers.

### 3.1  Evaluation Setup

All experiments were performed in a one-year-old computer with the configurations Intel(R) Core(TM)2 CPU 2.13GHz, 2 GB RAM, and the operation system platform is Microsoft Windows XP Professional (SP2). We have used an open source machine learning package – Weka (the latest Windows version: Weka 3.5.7). Weka is a collection of machine learning algorithms for data mining tasks that contains tools for data preprocessing, classification, regression, clustering, association rules, and visualization. This empirical study, however, only deals with a subset of classifier algorithms.

All the machine learning techniques that will be used in this paper are implemented in Weka so that they will be easily and fairly compared to each other.

The dataset to be used in our experiments in KDD99 labeled dataset. The main reason we use this dataset is that we need relevant data that can easily be shared with other researchers, allowing all kinds of techniques to be easily compared in the same baseline. The common practice in intrusion detection to claim good performance with "live data" makes it difficult to verify and improve pervious research results, as the traffic is never quantified or released for privacy concerns. The KDD99 dataset might have been criticized for its potential problems [13], but the fact is that it is the most widespread dataset that is used by many researchers and it is among the few comprehensive datasets that can be shared in intrusion detection nowadays.

As our test dataset, the KDD99 dataset contains one type of normal data and 24 different types of attacks that are broadly categorized in four groups of DoS, Probes, U2R and R2L. The packet information in the original TCP dump files were summarized into connections. This process is completed using the Bro IDS, resulting in 41 features for each connection (and one final feature for classifying, of course). Therefore, each instance of data consists of 41 features and each instance of them can be directly mapped into the point discussed in classifiers algorithms.

Due to the huge number of audit data records in the original KDD99 dataset as shown in Table 1, we have to manually random sample twice from the original KDD99 dataset. For the first time, we extracted 49,596 instances as training set for our experiments. They include 9,841 normal instances, 39,092 DoS instances, 437 Probe instances, 13 U2R instances and 213 R2L instances. Secondly, we extracted 15,437 instances as the independent testing set. By using these two datasets, we thus can effectively evaluate the performance of different methods.

**Table 1.** Distribution of connection types in KDD99 10% training dataset

| Class  | Number of records | % of occurrence |
|--------|-------------------|-----------------|
| Normal | 97,277            | 19.69           |
| DoS    | 391,458           | 79.24           |
| Probe  | 4,107             | 0.83            |
| U2R    | 52                | 0.01            |
| R2L    | 1,126             | 0.23            |
| Total  | 494,020           | 100.00          |

**Table 2.** Distribution of connection types in our actual training data for classifiers evaluation

| Class  | Number of records | % of occurrence |
|--------|-------------------|-----------------|
| Normal | 9,841             | 19.84           |
| DoS    | 39,092            | 78.82           |
| Probe  | 437               | 0.88            |
| U2R    | 13                | 0.03            |
| R2L    | 213               | 0.43            |
| Total  | 49,596            | 100.00          |

### 3.2   Classifier Algorithms

#### 3.2.1   BayesNet
*BayesNet* [8] learns Bayesian networks under the presumptions: nominal attributes (numeric one are predescretized) and no missing values (any such values are replaced globally). There are two different parts for estimating the conditional probability tables of the network. We run *BayesNet* with the SimpleEstimator and K2 search algorithm without using ADTree.

#### 3.2.2   NaïveBayes
The *NaïveBayes* [8] classifier provides a simple approach, with clear semantics, to representing and learning probabilitistic knowledge. It is termed *naïve* because is relies on two important simplifying assumes that the predictive attributes are conditionally independent given the class, and it posits that no hidden or latent attributes influence the prediction process.

#### 3.2.3   J48 (C4.5 Decision Tree Revision 8)
Perhaps *C4.5* algorithm which was developed by Quinlan [9] is the most popular tree classifier. Weka classifier package has its own version of *C4.5* known as *J48*. *J48* is an optimized implementation of *C4.5* rev. 8. *J48* is experimented is this study with the parameters: confidenceFactor = 0.25; numFolds = 3; seed = 1; unpruned = False.

#### 3.2.4   NBTree
*NBTree* [10] is a hybrid between decision trees and *NaïveBayes*. It creates trees whose leaves are *NaïveBayes* classifiers for the instances that reach the leaf. It is quite reasonable to expect that *NBTree* can outperform *NaïveBayes*; but instead, we may have to scarify some speed.

#### 3.2.5   Decision Table
*Decision Table* [3] builds a decision table majority classifier. It evaluates feature subsets using best-first search and can use cross-validation for evaluation. There is a set of methods that can be used in the search phase (E.g.: BestFirst, RankSearch, GeneticSearch …) and we may also use LBk to assist the result.  In this experiment, we choose the crossVal  =  1; searchMethod = BestFirst and useIBk = False

#### 3.2.6   JRip (RIPPER)

*RIPPER* [3] is one of the basic and most popular algorithms. Classes are examined in increasing size and an initial set of rules for the class is generated using incremental reduced-error pruning. We evaluate *RIPPER* through *JRip*, an implementation of *RIPPER* in Weka with the parameters: folds = 3; minNo = 2; optimizations = 2; seed = 1; usePruning = true.

#### 3.2.7   OneR
*OneR* [3] is another basic algorithm using Rule based model. It generates a one-level decision  tree expressed in the form of a set of rules that all test one particular attribute. *OneR* is a simple, cheap method that often comes up with quite good rules for characterizing the structure in data.

### 3.2.8 Multilayer Perceptron (MLP)

*Multilayer perceptron* (*MLP*) [11] is one of the most commonly used neural network classification algorithms. The architecture used for the *MLP* during simulations with KDD dataset consisted of a three layer feed-forward neural network: one input, one hidden, and one output layer. Selected parameters for the model are: learningRate = 0.3; momentum = 0.2; randomSeed = 0; validationThreshold = 20.

### 3.2.9 SMO

*SMO* [3] implements the sequential minimal optimization algorithm for training a support vector classifier, using polynomial or Gaussian kernels. *SMO* is evaluated with the following parameters: c = 1.0; epsilon = 1.0E-12; kernel = PolyKernel; numFolds = -1; randomSeed = 1.

### 3.2.10 LBk

*LBk* [12] is a lazy classifier algorithm that makes use of the k-nearest-neighbor classifier. In this study, we choose the parameters for *LBk* as follow: k = 1; crossValidate = False; searchAlgorithm = LinearNNSearch; windowSize = 0.

### 3.3 Performance Comparison

Best performing instances of all the ten classifiers selected in Section 3.2 were evaluated on the KDD dataset. Simulation results are given in the Table 4. To compare the classifiers, we record TP and FP of each algorithm. These parameters will be the most important criteria for the classifier to be consider the best algorithm for the given attack category. Besides, it is also at equal importance to record Average Accuracy (AA = Total correctly classified instances/Total instances) and Training Time (TT) of each algorithm. In the selection process, one algorithm will be disqualified if its AA is too low, despite its outstanding performance in one specific attack category. TT on the other hand, will give us the idea about which algorithm can be implemented in a real-time network intrusion detection system.

Just as we expected, Table 1 shows that no single algorithm could detect all attack categories with a high probability of detection and a low false alarm rate. It strengthen our belief that different algorithms should be used to deal with different types of network attacks. Results also show that for a given attack category, certain algorithms demonstrate superior detection performance compared to others. For DoS category, most algorithms yield very high TP rates – averagely 95%. *NaïveBayes* is the only one that lags behind as it gives a TP at 79.2%. But for Probe attacks, *NaïveBayes* outperforms the others with its FP at 94.8%; *BayesNet* and *Decision Table* both have impressive performance for this category at 83.8%. In U2R attacks, *BayesNet* and *Decision Table* are the best two classifiers with FP at 30.3% and 32.8% respectively.
And for the case of R2L attacks, only *OneR* could produce about 10% of attacks while the others just lag behind with inappreciable results.

**Table 4.** Performance comparison of the ten classifier algorithms – TP, FP and AA is measured in %. TT is measured in second.

| Classifier Category | Classifier Algorithm | | DoS | Probe | U2R | R2L |
|---|---|---|---|---|---|---|
| Bayes | BayesNet | TP | 94.6 | 83.8 | 30.3 | 5.2 |
| | | FP | 0.2 | 0.13 | 0.3 | 0.6 |
| | | AA | | 90.62 | | |
| | | TT | | 6.28 | | |
| | NaïveBayes | TP | 79.2 | 94.8 | 12.2 | 0.1 |
| | | FP | 1.7 | 13.3 | 0.9 | 0.3 |
| | | AA | | 78.32 | | |
| | | TT | | 5.57 | | |
| Trees | J48 | TP | 96.8 | 75.2 | 12.2 | 0.1 |
| | | FP | 1 | 0.2 | 0.1 | 0.5 |
| | | AA | | 92.06 | | |
| | | TT | | 15.85 | | |
| | NBTree | TP | 97.4 | 73.3 | 1.2 | 0.1 |
| | | FP | 1.2 | 1.1 | 0.1 | 0.5 |
| | | AA | | 92.28 | | |
| | | TT | | 295.88 | | |
| Rules | Decision Table | TP | 97.0 | 57.6 | 32.8 | 0.3 |
| | | FP | 10.7 | 0.4 | 0.3 | 0.1 |
| | | AA | | 91.66 | | |
| | | TT | | 66.24 | | |
| | JRip | TP | 97.4 | 83.8 | 12.8 | 0.1 |
| | | FP | 0.3 | 0.1 | 0.1 | 0.4 |
| | | AA | | 92.30 | | |
| | | TT | | 207.47 | | |
| | OneR | TP | 94.2 | 12.9 | 10.7 | 10.7 |
| | | FP | 6.8 | 0.1 | 2 | 0.1 |
| | | AA | | 89.31 | | |
| | | TT | | 3.75 | | |
| Functions | MLP | TP | 96.9 | 74.3 | 20.1 | 0.3 |
| | | FP | 1.4 | 0.1 | 0.1 | 0.5 |
| | | AA | | 92.03 | | |
| | | TT | | 350.15 | | |
| | SMO | TP | 96.4 | 74.3 | 13.3 | 0.1 |
| | | FP | 0.8 | 0.3 | 0.1 | 0.4 |
| | | AA | | 91.65 | | |
| | | TT | | 192.16 | | |
| Lazy | LBk | TP | 96.7 | 72.4 | 22.3 | 7.8 |
| | | FP | 0.8 | 0.2 | 0.1 | 0.6 |
| | | AA | | 92.22 | | |
| | | TT | | 10.63 | | |

## 4 Classifier Selection Model

After experiencing the performance comparison with the ten classifier algorithms, we generalize the empirical results with a model for algorithm selection. Observation from Table 4 suggests that for a given attack category, certain subset of classifier algorithms offer superior performance over the others. It is quite reasonable to expect much performance improvement as we can select the best classifier candidate for a specific attack category at hand. Section 3 identified the best algorithms for each attack categories: *JRip* for DoS and Probe, *Decision Table* for U2R and *OneR* for R2L. We then propose a model for classifier selection as in Fig. 1(a).

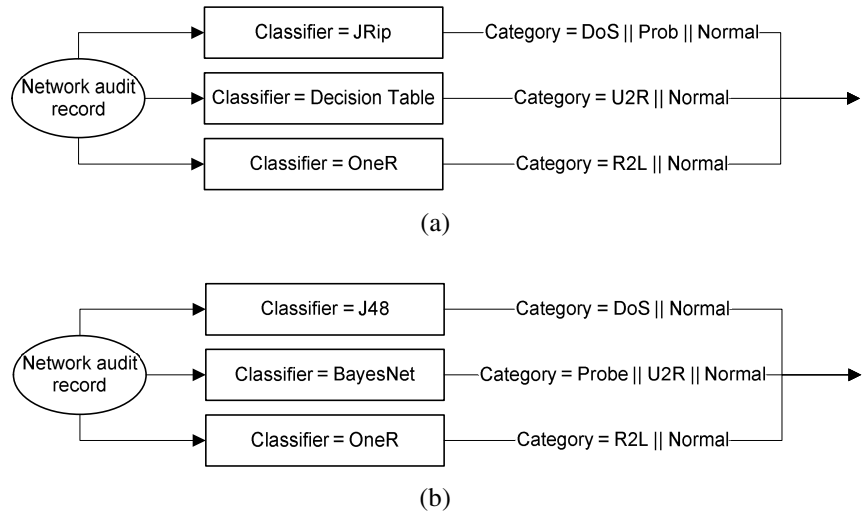

**Fig. 1. (a)**. Parallel model for classifier selection. **(b)**. Parallel model for real-time application classifier selection

About the application for the model proposed from the Fig. 1(a), it is expected that network intrusion detection systems with data mining capability should be flexible in choosing the classifying technique that best to deal with the attack at hand. Although detection rate improvement is something we expect, it is at equal importance to judge whether the selected algorithms can be implemented in real-time NIDs. The original need of network intrusion detection actually comes from the commercial world, not from the scientific community. Companies will pay more attention to classifier algorithms that can not only effectively detect network intrusion, but also can detect them in a really short time. Algorithms with high detection rate but end up consuming too much time in fact only are used by researchers in laboratories. Therefore, we suggest another model for real-time algorithm selection as in Fig. 1(b). The best algorithms with low TT for each attack category: *J48* for DoS, *BayesNet* for Probe and U2R and *OneR* for R2L. This model may have significant meaning for companies seeking for implementing real-time IDSs, or researchers that are currently working on developing light-weight data-mining algorithms.

## 5   Model Evaluation and Discussion

Table 5 shows the performance comparison of the two proposed multi-classifier models with KDD Cup Winner. The results suggest that the two proposed models showed minor improvement in TP for DoS and Probe; and significant improvement for U2R and R2L attack categories. Also, FP was reasonably small for all attack categories.

Table 5. Performance comparison between the two models and KDD Cup Winner

|  |  | DoS | Probe | U2R | R2L |
|---|---|---|---|---|---|
| KDD Cup Winner | TP | 97.10 | 83.30 | 13.20 | 8.40 |
|  | FP | 0.60 | 0.30 | 0.03 | 0.05 |
| Model 1(a) | TP | 97.40 | 83.80 | 32.80 | 10.70 |
|  | FP | 0.30 | 0.10 | 0.30 | 0.10 |
| Model 1(b) | TP | 96.80 | 83.80 | 30.30 | 10.70 |
|  | FP | 1.00 | 0.13 | 0.30 | 0.10 |

Despite the superior in numeric comparison between the proposed models and other approaches, one may have questions on some potential problems when the models are practically deployed in real systems: (1) Deployment of a system with multiple algorithms is inflexible since we may have to hardcode the algorithms. And if we have a better dataset than KDD99, the one used in this paper, then the best classifiers may be different and changing the classifiers will become such annoying. (2) Another problem when the models are implemented is the resource requirements. Consider ingress points of multi-gigabit networks, running multiple algorithms for intrusion detection at these locations may hurt the entire system's network performance. (3) And finally, a comparison between the proposed models and a multiple classifiers selection (MCS) system may be made.

In fact, the above mentioned issues may be well solved, one way or another. (1) To our best knowledge, KDD99 is the only dataset which is well tailored for data mining. There exist many other datasets for network intrusion detection (DRAPA, MAWI, NLANR …). But they are all "raw" network data, which will require a lot of process before they can be used for data mining applications. Suppose that we have another dataset which are more representative than KDD99 then the best classifiers may indeed be different. But if the system is flexibly designed (no such thing as "hardcode the models"), then modifying the classifier selection models will be easy. (2) Considering the resource problem, x3 or x4 processing time will not exist in reality since the models are configured in a parallel basis, which means that classifiers can be implemented by parallel processors to achieve real-time performance. (3) Finally, MCS systems are still currently designed in the labs. Researchers are proposing different systems. But there exists many problems and currently, none can achieve a stable system with optimistic performance yet. System using the proposed models will surely not be as dynamic as a MCS system, but it can provide stable and reliable performance.

## 6 Conclusion

For the contribution of this paper, firstly we made an up-to-date survey on recent studies about network intrusion detection that was evaluated with KDD99 dataset. We then use Weka to bring out an extensive performance comparison among the most popular classifier algorithms. Finally, two models for algorithm selection are proposed with great promise for performance improvement and real-time systems application. However, we are fully aware of problems that have been cited with the KDD99 dataset [13] and strongly discourage its further use in developing network intrusion detection data mining algorithms. The reason we use KDD99 in this paper, as explained, is because we need a baseline to evaluate different algorithms and compare our work with the others. For the future directions, we would like to evaluate our work on another dataset. Besides, we would also like to make real implementations of our algorithm selection models to practically experiment its effectiveness.

## Acknowledgement

This research was supported by the Industry Promotion Project for Regional Innovation. The authors would like to thank Prof. Park Hyukro and anonymous reviewers for their valuable comments and suggestions to this paper.